\documentclass[12pt,preprint]{aastex}

\newcommand{\beq}{\begin{equation}}
\newcommand{\eeq}{\end{equation}}
\newcommand{\beqa}{\begin{eqnarray}}
\newcommand{\eeqa}{\end{eqnarray}}

\title{A Catalogue of Morphologically Classified Galaxies from the
Sloan Digital Sky Survey: North Equatorial Region }
\author{Masataka Fukugita$^{(a,b)}$, Osamu Nakamura$^{(c)}$, 
Sadanori Okamura$^{(d)}$, Naoki Yasuda$^{(a)}$,\\
John C. Barentine$^{(e)}$, Jon Brinkmann$^{(e)}$, 
James E. Gunn$^{(f)}$, Mike Harvanek$^{(e)}$,\\
Takashi Ichikawa$^{(g)}$, Robert H. Lupton$^{(f)}$, 
Donald P. Schneider$^{(h)}$, Michael A. Strauss$^{(e)}$, 
Donald G. York$^{(i)}$ } 
\affil{$^{(a)}$Institute for Cosmic Ray Research,
University of Tokyo,\\ Kashiwa 277 8582, Japan}
\affil{$^{(b)}$Institute for Advanced Study, Princeton, NJ 08540, USA}
\affil{$^{(c)}$Graduate School of Political Science, Waseda
University, Shinjiku, Tokyo 169 8050, Japan} 
\affil{$^{(d)}$Department of Astronomy, University of Tokyo, Hongo, 
Tokyo 113, Japan}
\affil{$^{(e)}$Apache Point Observatory, Sunspot, NM 88349, USA}
\affil{$^{(f)}$Princeton University Observatory, Princeton University,
Princeton, NJ 08544, U. S. A.}  
 \affil{$^{(g)}$Astronomical
Institute, Tohoku University, Sendai 980 8578, Japan}
\affil{$^{(h)}$Department of Astronomy and Astrophysics, Pennsylvania
State University. University Park, PA 6802, USA}
\affil{$^{(i)}$Department of Astronomy and Astrophysics, University of
Chicago, Chicago, IL 60637, USA}
\begin{document}

\begin{abstract}

We present a catalogue of morphologically classified bright galaxies
in the north equatorial stripe (230 deg$^2$) derived from the Third
Data Release of the Sloan Digital Sky Survey (SDSS). Morphological
classification is performed by visual inspection of images in the $g$
band. The catalogue contains 2253 galaxies complete to a magnitude
limit of $r=16$ after Galactic extinction correction, selected from
2658 objects that are judged as extended in the photometric catalogue
in the same magnitude limit.  1866 galaxies in our catalogue have
spectroscopic information. A brief statistical analysis is presented
for the frequency of morphological types and mean colours in the
catalogue.  A visual inspection of the images reveals that the rate of
interacting galaxies in the local Universe is approximately 1.5\%
in the $r\le16$ sample.
A verification is made for the photometric catalogue generated by the
SDSS, especially as to its bright end completeness.

\end{abstract}


\section{Introduction}

This paper presents a catalogue of morphologically classified galaxies
from the Sloan Digital Sky Survey (SDSS: York et al. 2000), Data
Release Three (DR3; Abazajian et al. 2005).
We limit our sample to a rectangular region of the equatorial area
in the Northern sky (R.A.$\approx 9^h.7-15^h.7$) of 230 square degrees
that comprises 2658 objects brighter than the $r$ band Petrosian
magnitude $r_P\leq 16$ that are listed as extended in the DR3
photometric catalogue. The classification is performed by visual
inspection by three people independently, and the final classification
is obtained from the mean. We obtain a sample of 2253 galaxies, of
which 1866 have spectroscopic information in the SDSS.

Visual classification is a labourious and somewhat subjective
procedure.  Nevertheless, this remains the best approach to classify
each galaxy into a Hubble type with high confidence, at least for
bright galaxies. There are a number of methods that use 
photometric and/or spectroscopic parameters developed for large scale
samples to classify galaxies. Those classifications are correlated
reasonably well with visual Hubble types, but are substantially
contaminated by galaxies that belong to obviously wrong classes if
visual inspection is made.  The identification of Sa galaxies is
particularly subtle. The classification of Sa galaxies often scatters
across elliptical to late spiral galaxies, if one uses photometric
and/or spectroscopic parameters as indicators.  On the other hand, it
is not difficult to identify disks of Sa galaxies with visual
inspection. Colour or spectroscopic parameters are sensitive to the
star formation activity, so galaxies that show such activities are
typically classified as a late type if those parameters would be used.
For the moment it is difficult to replace the classification with a
quantitative measure.

Our work follows the traditional line of the Revised Shapley Ames
Catalogue (Sandage \& Tammann 1980), the Reference Catalogue of Bright
Galaxies (de Vaucouleurs \& de Vaucouleurs 1964, RC1; de Vaucouleurs et
al. 1995, RC3), the Uppsala General Catalogue of Galaxies (Nilson 1973) and
several others (e.g., Marzke et al. 1994; Kochanek et al. 2001; see
also Blanton et al. 2005; Driver et al. 2006, which carried out
rudimentary visual classifications).  The size of our sample is
moderate, but it is based on accurate photometric criteria to define
the basic catalogue and provides a photometrically homogeneous sample
that can be used for a variety of galaxy studies. We have
endeavoured to alleviate subjectivity of visual classification by
taking a mean of three independent classifiers.

This catalogue also allows us to verify the quality of the SDSS
photometric catalogue at the bright end. Definite photometric criteria
are applied to produce a galaxy sample that is to be targeted for
spectroscopic observations (Strauss et al. 2002). The problem is that
we are not able to produce a genuine galaxy catalogue with simple
numerical criteria.  The catalogue obtained may thus contain objects
that are not genuine galaxies, such as double stars, stars with
somewhat deformed images, ghosts, satellite images and `shredded'
objects (caused by failures in deblending of large bright galaxies).
On the contrary, an application of stricter criteria would miss many
true galaxies, so that a compromise is needed. We visually inspect all
objects that are selected with a rather loose criterion for extended
objects, which permits quantification of the contamination and
completeness of the photometric catalogue produced by the SDSS target
selection on the bright end.

We also give a subsidiary catalogue of $r'_P\le 15.9$ galaxies that
were contained in the Early Data Release (EDR: Stoughton et al. 2002),
since a number of scientific publications
(Nakamura et al. 2003; Ohama 2003 (TBC:2002?);
Fukugita et al. 2004; Ball et al. 2004; Nakamura et al. 2004;
Yamauchi et al. 2005; Tasca \& White 2005)
have used this earlier 
version of morphologically classified galaxy cataloge.  
Identification of all objects in both
catalogues is also given.
We note that the revision in the estimate of morphological
type is small, if any, for individual galaxies and
the results given in the earlier papers will
change little with the use of the present catalogue.

We refer the reader to the other publications for detailed
descriptions of the SDSS related to our study: Gunn et al (2006) for
the telescope, Gunn et al. (1998) for the photometric camera, Fukugita
et al. (1996) for the photometric system, Hogg et al. (2001) and Smith
et al. (2002) for external photometric calibrations, and Pier et al.
(2003) for astrometric calibrations.  We also refer to Abazajian et
al. (2003; 2004) and Adelman-McCarthy et al. (2006) for other data
releases from the SDSS, which discuss the successive improvement of
the pipelines used to derive the basic catalogues.

\section{Procedures}

Our rectangular region is defined by $145^\circ<\alpha_{\rm
J2000}<236^\circ$ and $-1^\circ.26<\delta_{\rm J2000}<1^\circ.26$,
covering an area of 230 deg$^2$.  This area fully encompasses SDSS
survey stripe 10 but the region of primary observations that takes
account of overlaps between stripes is somewhat rounded towards its
corners. So we supplement the missed part from neighbouring stripes
(stripes 9 and 11) to make the area strictly rectangular in celestrial
coordinates. We take photometry from DR3, and select all extended
objects\footnote{Technically, the selection is made using flag {\tt
type}=3 (galaxy), not {\tt saturated}, and not {\tt satur\_center} for
all colour bands.  Objects that are flagged as {\tt type}=3, and not
{\tt satur\_center} but flagged as {\tt saturated} are visually
inspected and included into the catalog when appropriate.  } that are
brighter than Petrosian magnitude $r_P=16$ (see EDR and Strauss et al. 
2002 for the precise definition) after the Galactic extinction
correction (Schlegel et al. 1998). There are 2658 such objects in the
DR3 catalogue that are produced from Version 5.4 of the photometric
pipeline.

We note that there are some gaps (0.03 deg$^2$ altogether) within the
region that concerns us. There are five fields ($13.'5\times 9.'0$)
for which the photometric pipeline did not process the data (the
actual gap is somewhat smaller due to overlaps with adjacent
fields). This happens when the field contains very bright galaxies or
stars with conspicuous spikes, and the completion of deblending
required more time than the pipeline limit.  Any galaxies located in
these fields are not included in our sample.

All objects are visually inspected by three classifiers (MF, ON, SO)
independently.  This sample is contaminated by non-galactic objects.
Our initial sample of 2658 objects includes a number of sources that
are not galaxies (stars, satellite trails, optical defects) as well as
multiple entries for a single object (primarily due to deblending
failures).  Removal of these objects produces a final sample of 2253
galaxies.

Morphological classification is carried out in reference to {\it The
Hubble Atlas of Galaxies} by Sandage (1961; Hubble Atlas hereafter) by
three classifiers using the SAOimage viewer. We use the monochromatic
$g$ band image, which is similar to the commonly-used $B$ band image
for classification, and is sensitive to HII regions and arm
structures.  It is important to inspect images with both linear and
logarithmic scales in the viewer with varying contrast levels. This
occasionally produces a systematic difference from classifications
based on photographic materials.  We intentionally avoid using colour
information so that morphology is solely determined by the
appearance, as has been done traditionally. This allows us to
consider the correlation between morphology and colours in an
unbiased way. 

We consider classification into 7 classes, $T=0$ (E), 1 (S0), 2(Sa),
3(Sb), 4(Sc), 5(Sd) and 6(Im), allowing for half-integer classes. We
do not adopt more detailed classes such as those defined in RC3 (which
has 16 classes in $T$), since our experience with the SDSS data
(comparing results from the three classifiers) is that a finer
division is unwarranted. The Hubble Atlas does not define Sd and Sdm.
We assign the latest of the Sc galaxies in the Hubble Atlas as
Sd$-$Sdm so that our classification scheme matches with that in
RC3. Irr I of the Hubble Atlas is denoted as Im in this paper, also in
agreement with RC3.  We assign $T=-1$ when we are unable to classify a
galaxy into a conventional Hubble type. We indicate by a symbol ``p''
when some peculiarity is noted in a galaxy (such as somewhat disturbed
shapes, rings, dust lanes etc. as in the Hubble Atlas)
even though it can be classified into a regular, $T\ne -1$, Hubble type.  The
relation between our $T$ and $T({\rm RC3})$ is shown in Table 1.  We
note slight biases in our classification towards integer $T$. We are
not concerned with whether galaxies have bars or not.

During the classification we noticed that not all galaxies fell
nicely into the Hubble sequence, but whenever reasonable
we classified a galaxy into the $T=0$ to 6 scale. This leads to
a number of cases where the appearance of the galaxy may not match well
with the appearance of the Hubble Atlas prototype, but we view this
as preferable to a catalogue containing a large number of 
unclassifiable ($T=-1$) objects. Our catalogue contains 33 galaxies
with $T=-1$.

We noted that some galaxies, notably among those with $T=-1$, show a
common feature, characterised by a high surface brightness and a
smooth light distribution, but their appearance is definitely not that
of E or S0 galaxies. These objects frequently have more irregular
shapes than early-type galaxies, but the light distribution is too
smooth and/or surface brightness too high to be classified as Im. They
appear to be reminiscent of Irr II in the Hubble Atlas or `amorphous'
galaxies introduced by Sandage and Brucato (1979), who characterised
them as `not E, S0, or any type of spiral no matter how peculiar, but
rather have an amorphous appearance to the unresolved
light'. Gallagher \& Hunter (1987) used the term amorphous to
represent `all galaxies with E/S0-like morphologies whose other global
properties resemble irregular galaxies.'  These definitions, however,
are not quite clear.  These galaxies tend to show a red colour. We
confirmed that most of the galaxies of this type in our sample 
show red colours, although we did not use the colour
information when we classified galaxies. We indicate by ``am'' when we
encounter those galaxies, whether they are left unclassified or
classified into regular types (see a figure given at the end of
Section 3).

If the SDSS photometric pipeline determines that an image is actually
composed of more than one component, the `parent' image is deblended into
`child' images (Lupton 2006). The child images may be
further deblended if they are judged to be formed of more than one
component.  When confronted with a large, complex surface brightness
pattern, the deblender `shreds' a bright extended galaxy into a
multitude of components, which can obviously affect the morphology of
the galaxy (and also photometric measurements).  (An extreme example
is the separation of a nucleus, which may look as S0, from a spiral
galaxy.)  For this reason it is important to inspect both parent and
child images for all objects to ensure correct classifications.

Each classifier independently carries out the classification of each
galaxy at least twice. The results are then compared, and when the
results for an individual galaxy differ by more than 1.5 units in $T$,
all classifiers reinspect the galaxy in question and make a final,
independent assessment.  The adopted morphological classification is
the mean of the measurements of the three classifiers.

The panels in Figure 1 display the correlation in $T$ among the three
classifiers.  The dispersion is 0.4, or $\approx$1 in the RC3 $T$
scale. This is probably as good as can be expected for visual
classification; for example, $\delta T({\rm RC3})=1.8$ in Lahav et al. 
(1995), a study that used photographic prints.

The final SDSS sample contains 218 galaxies that have assigned
$T$(RC3) in the RC3 catalogue. Figure 2 shows the correlation of
T(RC3) versus those from our classification for those common galaxies.
The correlation is generally good; however, there is a systematic
difference in the classifications in that S0/a to Sa(Sab) galaxies in
RC3 are classified somewhat later to Sa to Sb (occasionally to Sc) in
our catalogue.  On the other hand, our E and S0 galaxy samples do not
contain any galaxies that are classified as S0/a or later in RC3.  We
suspect the main reason of the discrepancy to be that our
classification is based on high dynamic range and high contrast CCD
images, which allows detection of arm structure and detailed texture
that could not be apparent in a single photometric image. This will
drive our classification of disc galaxies towards later types,
compared to those in the RC3 catalogues.

The quality of the photometry is also examined by visual inspection of
images, to check whether the SDSS atlas image contains the entire
image of the galaxy. If these data contain extra objects or parts of
the galaxy are erroneously removed by the deblender, flags are
attached. The position of spectroscopic fibre, which has a diameter of
$3''$, is also inspected.  When the fibre is not centred on the
nucleus, but is assigned to a region of the galaxy (e.g., a bright HII
region), the spectroscopic information is accepted but flagged.

\section{Catalogue}

Table 2 presents our final catalogue from DR3 ($r_P\le 16$) in the
order of right ascension, but only the top 20 lines are shown in the
printed version of the paper. The entire catalogue is available in an
electronically readable format.  The catalogue contains 15 columns
with the following information:

\begin{itemize}
\item
Column 1: 
catalogue number;
\item
Column 2: 
photometric identification number in DR3. The numbers
mean run (observation) $-$ rerun (photometric
data reduction) $-$ camcol (camera column) $-$ field $-$ object id
(see DR3 for details);
\item
Column 3: 
right ascension (J2000), in decimal degrees;
\item
Column 4: 
declination (J2000), in decimal degrees;
\item
Column 5: 
photometric identification number in EDR (see explanation for Column 2);
\item
Column 6: 
$I_{\rm sample}$ flag for catalogue inclusions: [3] 
in both DR3 ($r_P\le 16$) and
EDR ($r_P\le 15.9$); [2] only in DR3 ($r_P\le 16$);
\item
Column 7: 
$I_{\rm target}$ flag for spectroscopic target selection: [0] not targeted;
[1] targeted but not observed; [2] observed; [2:] fibre positioned off-nucleus,
but on some other part of the object carrying the
specified photometric identification;
\item
Column 8: 
morphological index $T$: the mean of three classifiers
rounded to the nearest integer or half-integer;
\item
Column 9: 
standard deviation of morphological indices among three classifiers;
\item
Column 10: 
$I_{\rm ph}$ photometry quality: [0] good photometry
(typical error expected to be smaller than approximately 0.1 mag based
on visual estimates); [1] photometry is accurate if one uses the
magnitude given to the parent image; [2] some errors, say 0.3 mag
(with visual estimates), is suspected in photometry; [3] poor
photometry. (Note that these flags are somewhat subjective.) Flags 2
and 3 are occasionally appended by p or c, which means that more
accurate magnitudes may easily be obtained by applying aperture
photometry centred on the designated object using parent or child
atlas image frames, respectively.
\item
Column 11: 
Petrosian $r$ magnitude (of the child image) after correcting for 
Galactic extinction;
\item
Column 12: 
spectroscopic identification number: 
spectroscopic plate number$-$ mjd $-$fibre number;
\item
Column 13: 
heliocentric redshift;
\item
Column 14: 
confidence level of redshift measurements;
\item
Column 15: 
remarks: `p' stands for peculiar, and this flag is
given only when galaxies are classified into normal types. `am' is
given to galaxies with an amorphous appearance. `int' stands for
interacting, and 'd-nucl' for double nuclei within a single galaxy.
`double' and `multiple' stand for more than one galaxy in the child
field, while interactions among them are not apparent.  The PGC number
is provided when the galaxy is identified with that listed in RC3.

\end{itemize}

In the bottom of Table 2 we append a
similar catalogue for 22 objects that are included only in our EDR
sample ($r_P\le 15.9$). All of the galaxies have Galactic-extinction
corrected $r_P>16$ in DR3; This change from the EDR measurement is due
to reprocessing the EDR data with the improved DR3 photometric
pipeline.  Flag [1] is assigned to Column 6, `flag for catalogue
inclusions'.  We carried out reclassifications for the EDR sample, but
the change compared to the earlier catalogue is insignificant.

Figures 3$-$7 display examples of $gri$ colour synthetic images of
galaxies (20 each) that are classified as $T=0-4$, taken from Catalog
Archive Server (CAS).  Figure 8 shows images for $T=5$ (8 galaxies in
the top) and $T=6$ (12 galaxies in the bottom).  Note that detailed textures
are not quite visible on these pictures and the contrast is not always
well represented, so that these printed images are not always appropriate
for the purpose of classification.  Figure 9 shows 12 galaxies, which
we classified as interacting; the four galaxies in the bottom row have
double (multiple) nuclei.  Figure 10 shows 16 galaxies to which we
give amorphous (``am'') flags.  The size of pictures are all $1'\times
1'$.

\section{Verification of the photometric catalogue of DR3}

We have adopted a set of inclusive selection criteria so that few
galaxies are missed in our initial sample.  This selection is
substantially looser than that adopted in the operational
spectroscopic target selection for galaxies (Strauss et al. 2002).

Among 2658 extended objects in our sample, 2253 are unique
galaxies. There are 27 examples of galaxies that are included two or
more times in the initial list; this primarily arises from deblending
difficulties.  Also included in our original set are 211 stars
(approximately 80\% of which are double stars) and 167 spurious
objects, such as satellite trails, diffraction spikes of bright stars,
ghosts, failures of deblending or of removal of bright stars that
saturate the CCD, and empty fields with no designated object (which
happen typically when imaging of the same fields with other colour
bands was hit by satellite trails). Nearly all stars (206 out of 211)
can be rejected if one imposes the condition $g({\rm PSF})-g({\rm
model})>0.5$ for the selection of galaxies, which is tighter
than the one used in target selection, $r({\rm PSF})-r({\rm
model})>0.3$. The former condition rejects six true galaxies (one
among them looks like an AGN); 61 spurious objects, however, escape the
rejection and contaminate the galaxy sample.

Among the 2253 galaxies, 2213 (98.2 \%) are chosen by SDSS target
selection, and 1866 (82.8\% of the entire galaxy sample) are actually
spectroscopically observed\footnote{Note that stripe 10 was observed
in early days of the SDSS observation, when the tuning of
spectroscopic observations was still immature. We suspect a higher
rate of spectroscopic observations for stripes observed in later
times.}.
The completeness of the spectroscopic observation in our sample
is essentially uniform from early- to late-types within Poissonian error.
One reason for missing spectroscopy is the fibre separation
constraint (fibres on a given plate must be separated by at least
$55''$, Blanton et al. 2003b). In cases where a galaxy and a quasar
candidate conflict, the fibre is assigned to the latter.  We found a
few patches for which spectroscopic observations were not carried out
for unknown reasons. We also found that 168 more targets are set by
target selection on non-galactic objects, of which 18 are
observed\footnote{We suspect that these rather large numbers are
likely due to deblending errors of the early immature version of
photometric pipeline used for spectroscopic target selection, since
spectroscopic observations are carried out in early stages of
SDSS operations for the region that concerns us in this paper.}

The survey samples are summarised in Table 3. The spectroscopic
sample is quite clean even for bright galaxies of our smaple, 
but with a completeness of
83\%. We wish to inject a note of caution to the use of SDSS
photometric galaxy catalogues. The target-selection algorithm of
Strauss et al. yields a sample of galaxies with good completeness,
only 2\% of galaxies missed, but suffers a 7\% contamination by stars
and spurious objects.

In order to examine the statistical completeness, we show in Figure 11
the number of galaxies as a function of $r$ magnitude.  The solid line
shows $N\sim 10^{0.6r}$ expected for Euclidean geometry. The data
indicate that the galaxy number count deviates little from this line
from 10 to 16 mag. A slight excess at bright magnitudes is consistent
with the Poisson statistics.  This implies that we have not missed too
many galaxies even in the bright end of the sample at 10$-$10.5 mag.
The spectroscopic sample is indicated by shading, which shows that the
spectroscopic completeness stays nearly constant for $r>12.5$. The
thick shades indicates galaxies that carry a flag for
photometric errors, which increases towards brighter magnitudes, 
from 5\% at $r=15.5$ to 20\% at $r=13$.
The number count for each morphological type
is shown in Figure 12. All curves are consistent with 
$N\sim 10^{0.6r}$ up to statistical errors, indicating
homogeneity in morphological compositions, and therefore
morphological fractions change little as a function of
magnitude to $r=16$.
We note, however, that the region considered has some
over-density at $z\sim0.8$ where more early-type
galaxies are included (see Figure 2 of Nakamura et al.). This may cause a
slight deviation from the smooth $10^{0.6r}$ growth.

An additional test is carried out for the completeness
by comparing galaxies in our sample
with those in RC3 and Updated Zwicky
Catalogues (Falco et al. 1999). The RC3 contains 269 galaxies in our
survey area; 15 of these are not in our catalogue. Nine of the 15 are
too faint to us ($r>16$), and 2 are omitted because they lie too close
to the edge of our area.  Three (PGC33550, PGC39695; PGC39705) are in
the field for which the SDSS photometric pipeline could not process
the frame due to the presence of the excessively bright stars (for the
first two) or the bright galaxy itself (PG33550, $B_T=9.8$). The one
(PGC53499=NGC5792) is a bright galaxy but lies too close to a very
bright star.  In summary, only 4 of 258 RC3 galaxies that should have
been included in our catalogue were missed.

A similar result was obtained in a comparison of the 394 updated
Zwicky Catalogue objects in our survey area; 14 of these objects were
missed.  One Zwicky galaxy (one of a pair of interacting galaxies) was
shredded by the deblender into components that all had $r>16$, and
hence dropped from our catalogue. In total, four bright Zwicky
galaxies (there are three in common to those we found for the case
of RC3) that should have been in our catalogue were missed by the SDSS
photometric pipeline.

From these tests we conclude that galaxies are well sampled to as
bright as 10 mag, unless they are accidentally located close to very
bright stars. The most important cause of missing bright galaxies is
failures of deblending in the presence of vary bright
stars or galaxies themselves; we missed the fraction of the region,
$\approx1.3\times10^{-4}$.  We expect that the incompleteness will
become an important issue for $r \lesssim 10$.
A comparison with the RC3 catalogue (which includes
all Zwicky galaxies) shows that
incompleteness for low-surface brightness galaxies
is no more than that in RC3.

\section{Statistics} 

Figure 13 shows histograms of the morphological type
distribution of
galaxies for both photometric and spectroscopic samples. We use only
seven classes, grouping half those classified into half-integer $T$
into each adjoining integer bins. The fractional morphological
composition of our catalogue breaks down to E: E/S0$-$S0: S0a$-$Sab:
Sb$-$Sc: Scd$-$Sdm: Im = 0.14: 0.26: 0.25: 0.28: 0.038: 0.014. A $B$
band study summarised by Fukugita, Hogan \& Peebles (1998) gives a
relative frequency of E: S0: Sab: Sbc: Scd: Im = 0.11: 0.21: 0.28:
0.29: 0.045: 0.061. 
A somewhat higher fraction of early type galaxies in our sample
is ascribed to our galaxy selection in the $r$ magnitude, which
will select a larger fraction of early-type galaxies than would be
present in a volume-limited sample. Our small fraction of Im galaxies
arises from the intrinsical small luminosity of these sources that
makes the sampling volume small.  These issues are discussed in
Nakamura et al. (2003), where the morphologically classified
luminosity function is derived.

We identified 25 galaxies which are interacting, and an additional six
that display features that suggest interacting.  Of this set of 31, 16
have such disturbed morphologies that they are assigned $T=-1$
(unclassified). In our galaxy sample 12 galaxies have double 
(or multiple) nuclei, and four of these are also counted as
``interacting'' and 1 suspected interacting. Adding double-nucleus
galaxies, we arrive at 33$-$38 interacting galaxies in our catalogue,
i.e., the rate of interacting galaxies in a nearby magnitude-limited
galaxy sample is 1.5-1.7\%.

The mean colours of galaxies after K corrections (Blanton et
al. 2003a) are given in Table 4. We have rejected galaxies for which
poor photometry is suspected ($I_{\rm ph}\ge 2$).  This information
supersedes the mean colours given in Shimasaku et
al. (2001)\footnote{Shimasaku et al. (2001) do not include K
corrections due to a lack of redshift at that time. Although the
galaxies are all at low redshift, the absence of K corrections make
galaxies, especially those of early types, redder by a non-negligible
amount.  Without the K correction the colours quoted below in this
paragraph will be 1.85, 0.89, 0.41 and 0.28 for the sample used in the
present paper.}.  The colours, except for $i-z$, form monotonic
sequences from red to blue with increasing $T$, including half integer
types that are not shown in this table.  The scatter of colours among
different galaxies at given $T$ is larger than the difference between
the mean colours of the neighbouring types. For example, the mean
colours of E galaxies for $u-g$, $g-r$ and $r-i$ are within one
standard deviation of those of Sa galaxies. The $i-z$ colours stay
essentially constant from E to Sab. For later types, some bluing trend
is present in $i-z$, but the scatter widens and is larger than the
variation.
The mean colours of E galaxies are $u-g=1.73\pm0.18$ (1.99),
$g-r=0.77\pm0.04$ (0.77), $r-i=0.39\pm0.03$ (0.43), $i-z=0.18\pm0.04$
(0.36), where the numbers in parentheses are the spectrosynthetic
calculation of Fukugita et al. (1995). There is a significant
disagreement in the reddest colours, as was noted by Shimasaku et al.
(2001).

\acknowledgments

This work was supported in Japan by Grant-in-Aid of the Ministry of 
Education. MF received support from the Monell Foundation at the 
Institute for Advanced Study.

Funding for the SDSS and SDSS-II has been provided by the Alfred P. Sloan 
Foundation, the Participating Institutions, the National Science Foundation, 
the U.S. Department of Energy, the National Aeronautics and Space 
Administration, the Japanese Monbukagakusho, the Max Planck Society, and 
the Higher Education Funding Council for England. The SDSS Web Site is 
http://www.sdss.org/.
The SDSS is managed by the Astrophysical Research Consortium for the 
Participating Institutions. The Participating Institutions are the 
American Museum of Natural History, Astrophysical Institute Potsdam, 
University of Basel, Cambridge University, Case Western Reserve University, 
University of Chicago, Drexel University, Fermilab, the Institute for 
Advanced Study, the Japan Participation Group, Johns Hopkins University, 
the Joint Institute for Nuclear Astrophysics, the Kavli Institute for 
Particle Astrophysics and Cosmology, the Korean Scientist Group, the 
Chinese Academy of Sciences (LAMOST), Los Alamos National Laboratory, 
the Max-Planck-Institute for Astronomy (MPIA), the Max-Planck-Institute 
for Astrophysics (MPA), New Mexico State University, Ohio State University, 
University of Pittsburgh, University of Portsmouth, Princeton University, 
the United States Naval Observatory, and the University of Washington.

\newpage

\begin{deluxetable}{lcccccccc}  
\tablecolumns{7}  
\tablewidth{0pc}  
\tablecaption{Morphological index $T$}
\tablehead{
\colhead{Hubble type} & \colhead{E} & \colhead{S0} & \colhead{Sa} &\colhead{Sb}&\colhead{Sc}&\colhead{Sd}&\colhead{Im}&\colhead{unclass.}}
\startdata
$T$(ours)&0&1&2&3&4&5&6&$-$1\cr
$T$(RC3)&$-$6 to $-$4&$-$3 to $-$1&1&3&5&7 to 8&10& \cr
\enddata
\label{table:counts}
\end{deluxetable}

\begin{deluxetable}{rcccccccccccccl}
\tabletypesize{\scriptsize}
\rotate
\tablecaption{Catalogue of Morphologically Classified Galaxies \label{tab:cat}}
\tablewidth{0pt}
\tablehead{
\colhead{ID} & 
\colhead{DR3 photo-ID} & 
\colhead{$\alpha_{\rm J2000}(^\circ)$} & 
\colhead{$\delta_{\rm J2000}(^\circ)$} & 
\colhead{EDR photo-ID}  &
\colhead{$I_{\rm sample}$} & 
\colhead{$I_{\rm target}$} & 
\colhead{$T$} & 
\colhead{$\sigma(T)$} & 
\colhead{$I_{\rm ph}$} & 
\colhead{$r_P$} & 
\colhead{Spectro-ID}&
\colhead{$z$}&
\colhead{CL($z$)}&
\colhead{remarks}
}
\startdata
1&0756-44-6-0195-0158& 145.00014 & 1.10623 &  - & 2 & 2 &  
0.5 & 0.4 & 0 & 15.64 & 477-52026-100 & 0.0605 & 0.999 \\[0.4em]         
2&0756-44-4-0195-0158& 145.04410 & 0.22011 &  - & 2 & 2 &  
3.0 & 0.0 & 0 & 15.49 & 476-52314-587 & 0.0621 & 0.998 \\[0.4em] 
3&0756-44-4-0195-0165& 145.04752 & 0.23774 &  - & 2 & 2 &  
1.0 & 0.4 & 0 & 15.25 & 476-52314-585 & 0.0622 & 0.999 \\[0.4em] 
4&0756-44-5-0195-0208& 145.06132 & 0.70924 &  -  & 2 & 2 &  
3.5 & 0.5 & 0 & 15.61 & 477-52026-98 & 0.0260 & 0.958 \\[0.4em]
5&0756-44-4-0196-0172& 145.22040 & 0.41082 &  756-4-8-0196-0174 & 3 & 2 &
3.5 & 0.6 & 0 & 15.89 & 266-51630-350 & 0.0982 & 1.000 \\[0.4em]
6&1239-40-1-0167-0166& 145.37412 &-1.24928 &  752-1-8-0012-0078 & 3 & 1 &  
0.0 & 0.4 & 0 & 14.89 &    -  & 0.000 & 0.000 \\[0.4em] 
7&0756-44-4-0198-0055& 145.51373 & 0.33644 &  - & 2 & 1 &  
4.0 & 0.0 & 2p & 11.77 &   -  & 0.000 & 0.000 & PGC2773\\[0.4em]
8&1239-40-2-0169-0142& 145.64788 & -0.77173&  752-2-8-0014-0175 & 3 & 2 &  
3.0 & 0.8 & 0 & 15.75 & 266-51630-215 & 0.0218 & 0.992 \\[0.4em]  
9&0756-44-1-0199-0259& 145.68110 & -0.86723 & 756-1-8-0199-0148 & 3 & 2 &  
2.0 & 0.5 & 0 & 15.60 & 266-51630-207 & 0.0676 & 0.999 \\[0.4em] 
10&1239-40-2-0170-0139& 145.75971 & -0.81389 & - & 2 & 2 &  
2.5 & 0.5 & 0 & 15.96 & 266-51630-216 & 0.0676 & 0.946 \\[0.4em]
11&1239-40-1-0170-0201& 145.76792 & -1.07472 & 752-1-8-0015-0167 & 3 & 1 &  
3.5 & 0.5 & 0 & 15.84 &     -  & 0.000 & 0.000 \\[0.4em]
12&0756-44-4-0200-0098& 145.80018 &  0.41417& 756-4-8-0200-0158 & 3 & 2 &  
4.0 & 0.4 & 0 & 14.18 & 266-51630-430 & 0.0252 & 0.996 & PGC27803\\[0.4em] 
13&0756-44-5-0200-0211& 145.84750 &  0.67573 & 0756-5-8-0200-0131 & 3 & 2 &  
2.5 & 0.5 & 0 & 15.87 & 266-51630-422 & 0.0266 &1.000 \\[0.4em] 
14&0756-44-6-0200-0100& 145.85049 &  1.20353 & 756-6-8-0200-0063 & 3 & 2 &  
1.5 & 0.5 & 0 & 15.60 & 480-51989-272 & 0.0618 & 1.000 \\[0.4em] 
15&1239-40-4-0170-0202& 145.87328 &  0.05683 & 752-4-8-0015-0058 & 3 & 1 &  
1.0 & 0.4 & 0 & 15.88 &     -   & 0.000 & 0.000 \\[0.4em]           
16&0756-44-2-0201-0130& 145.87445 & -0.60876 & 756-2-8-0201-0156 & 3 & 2 &  
4.0 & 0.0 & 0 & 15.89 & 266-51630-138 & 0.0715 & 0.999 \\[0.4em] 
17&0756-44-6-0201-0022& 145.89254 &  1.11773 & -  & 2 & 2 & 
-1.0 & 0.0 & 0 & 15.93 & 480-51989-266 & 0.0512 & 0.986 & int, am \\[0.4em]
18&1239-40-5-0171-0179& 145.94781 &  0.46530 & 752-5-8-0016-0090 & 3 & 2 &  
1.0 & 0.8 &  0 & 15.17 & 266-51630-467 & 0.0304 & 0.997 \\[0.4em]
19&1239-40-2-0171-0091& 146.00780 & -0.64227 & 752-2-8-0016-0100 & 3 & 2 & 
-1.0 & 0.0 & 2p & 15.89 & 266-51630-100 & 0.0051 & 0.938 &  \\[0.4em] 
20&0756-44-5-0202-0018& 146.02092 &  0.73355 & 756-5-8-0202-0009 & 3 & 2 &  
0.5 & 0.6 & 0 & 14.10 & 266-51630-461 & 0.0362 & 0.999 \\[0.4em]           
\enddata
\end{deluxetable}

\begin{deluxetable}{lcccc}  
\tablecolumns{7}  
\tablewidth{0pc}  
\tablecaption{Numbers of objects}
\tablehead{
\colhead{Galaxy sample} & \colhead{Galaxy} & \colhead{Double counted} & \colhead{Star/spurious} &\colhead{Initial sample}}
\startdata
Photometric sample......................... & 2253 & 27 & 378 & 2658 \cr
Targetted spectroscopic sample....... & 2213 & 20  & 168 & 2401 \cr
Observed spectroscopic sample......... & 1866 & 0  &  18 & 1884 \cr
\enddata
\label{table:counts}
\end{deluxetable}

\begin{deluxetable}{lcccccccc}  
\tablecolumns{7}  
\tablewidth{0pc}  
\tablecaption{Statistical properties of galaxies}
\tabletypesize{\scriptsize}
\tablehead{
\colhead{Hubble type} & \colhead{E} & \colhead{S0} & \colhead{Sa} &\colhead{Sb}&\colhead{Sc}&\colhead{Sd}&\colhead{Im}}
\startdata
Number & 265 & 255 & 139 & 188 & 166 & 9 & 18 \cr
$u-g$&1.73$\pm$0.18&1.65$\pm$0.21&1.50$\pm$0.29&1.33$\pm$0.28&1.35$\pm$0.26&
1.18$\pm$0.10&1.15$\pm$0.34\cr
$g-r$&0.77$\pm$0.04&0.74$\pm$0.07&0.68$\pm$0.10&0.60$\pm$0.13&0.54$\pm$0.10& 
0.47$\pm$0.09&0.36$\pm$0.13 \cr 
$r-i$&0.39$\pm$0.03&0.38$\pm$0.04&0.35$\pm$0.05&0.31$\pm$0.09&0.26$\pm$0.08&
0.16$\pm$0.08&0.09$\pm$0.11\cr
$i-z$&0.18$\pm$0.04&0.19$\pm$0.05&0.18$\pm$0.07&0.15$\pm$0.09&0.06$\pm$0.13&
0.01$\pm$0.15&$-$0.06$\pm$0.21\cr
\enddata
\tablecomments{Those galaxies that are classified as a half-integer type
are omitted from these statistics. The error stands for the dispersion. }
\label{table:counts}
\end{deluxetable}

\begin{figure}
\plotone{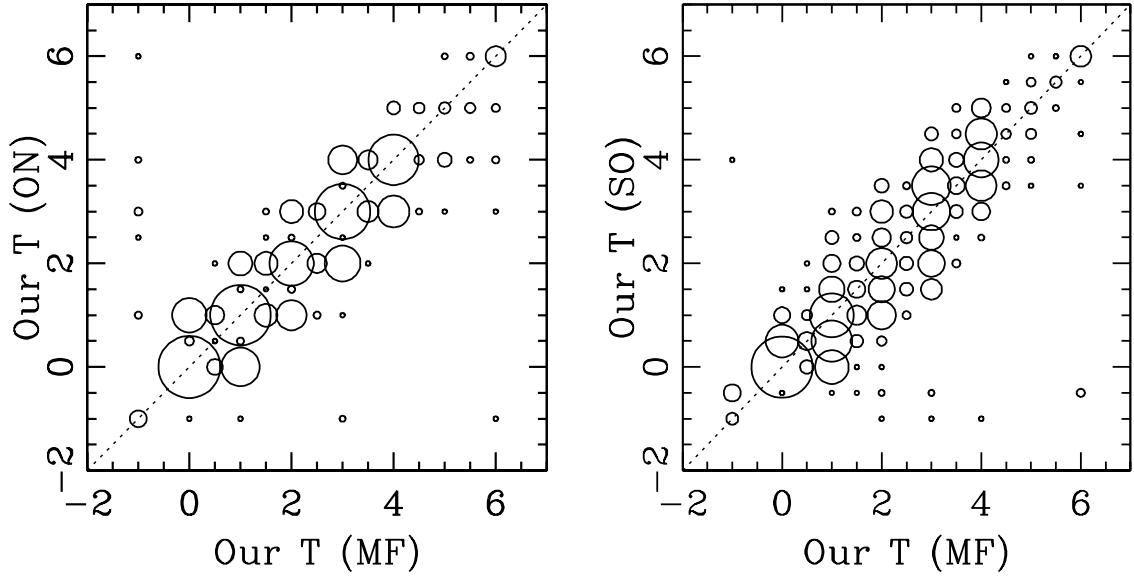}
\caption{Correlation of visually inferred morphological types
among three classifiers. The area of the circle represents the
number of galaxies in the grid. \label{fig1}}
\end{figure}

\begin{figure}
\plotone{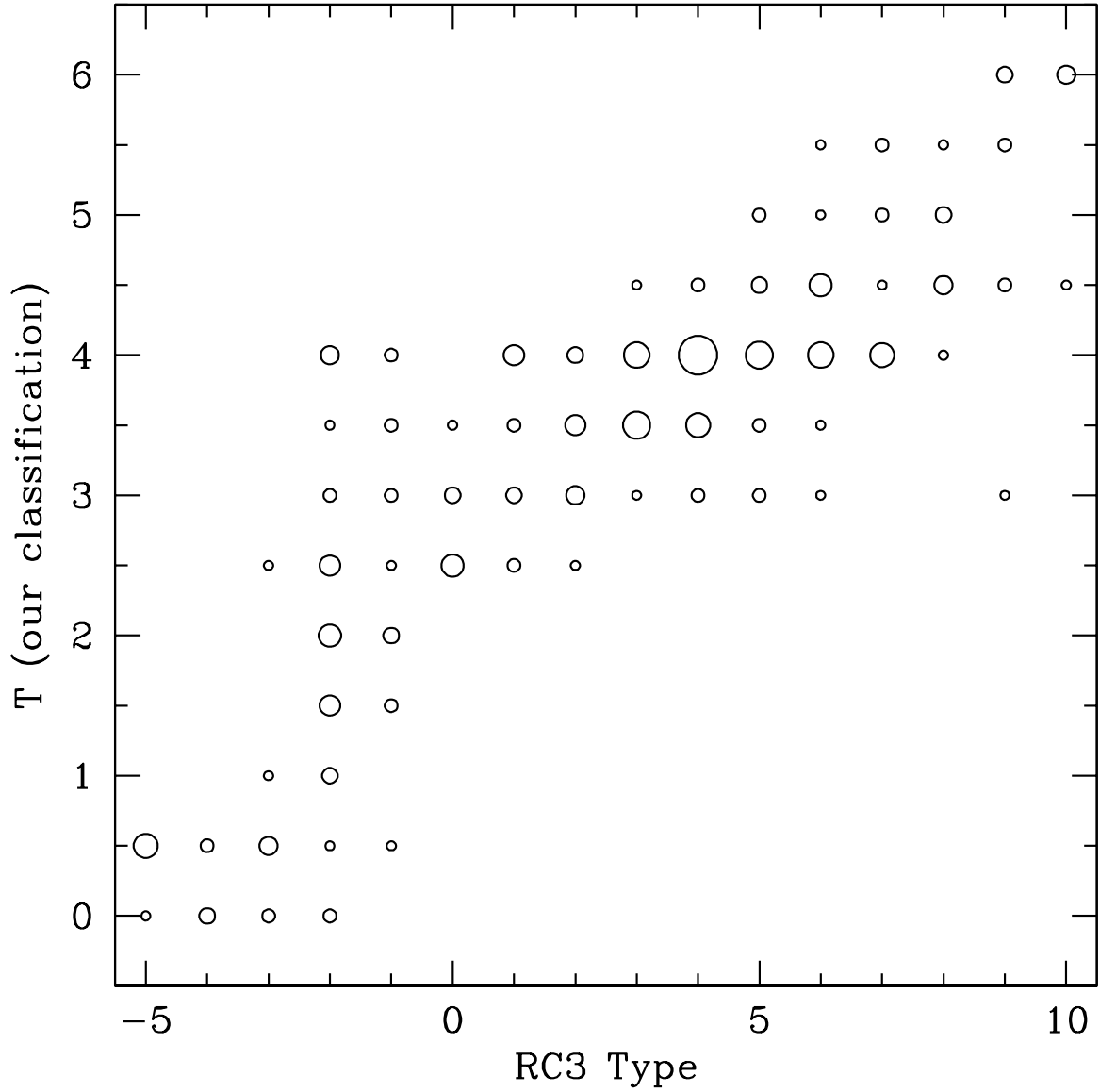}
\caption{Correlation of our $T$ with those of RC3 for 218 galaxies
common to the two samples. The area of the circle represents the
number of galaxies in the grid. \label{fig2}}
\end{figure}

\begin{figure}
\epsscale{0.9}
\caption{Sample of E ($T=0$) galaxies with synthetic $gri$ colour. The size is $1'\times 1'$ for all
pictures. \label{fig3}}
\end{figure}

\begin{figure}
\caption{Sample of S0 ($T=1$) galaxies. The format is the same as for
Figure \ref{fig3}. \label{fig4}}
\end{figure}

\begin{figure}
\caption{Sample of Sa ($T=2$) galaxies. The format is the same as for
Figure \ref{fig3}.\label{fig5}}
\end{figure}

\begin{figure}
\caption{Sample of Sb ($T=3$) galaxies. The format is the same as for
Figure \ref{fig3}.\label{fig6}}
\end{figure}

\begin{figure}
\caption{Sample of Sc ($T=4)$ galaxies. The format is the same as for
Figure \ref{fig3}.\label{fig7}}
\end{figure}

\begin{figure}
\caption{Sample of Sd ($T=5$) and Im ($T=6$) galaxies. The top two
rows display Sd galaxies, and the bottom three Im galaxies. The other
format is the same as for Figure \ref{fig3}.
\label{fig8}}
\end{figure}

\begin{figure}
\caption{Sample of interacting galaxies. The 4 galaxies in the bottom
row are galaxies with double nuclei. The format is the same as for
Figure \ref{fig3}.\label{fig9}}
\end{figure}

\begin{figure}
\caption{Sample of galaxies with ``amorphous'' appearance. The format
is the same as for Figure \ref{fig3}.
\label{fig10}}
\end{figure}

\begin{figure}
\epsscale{1.0}
\plotone{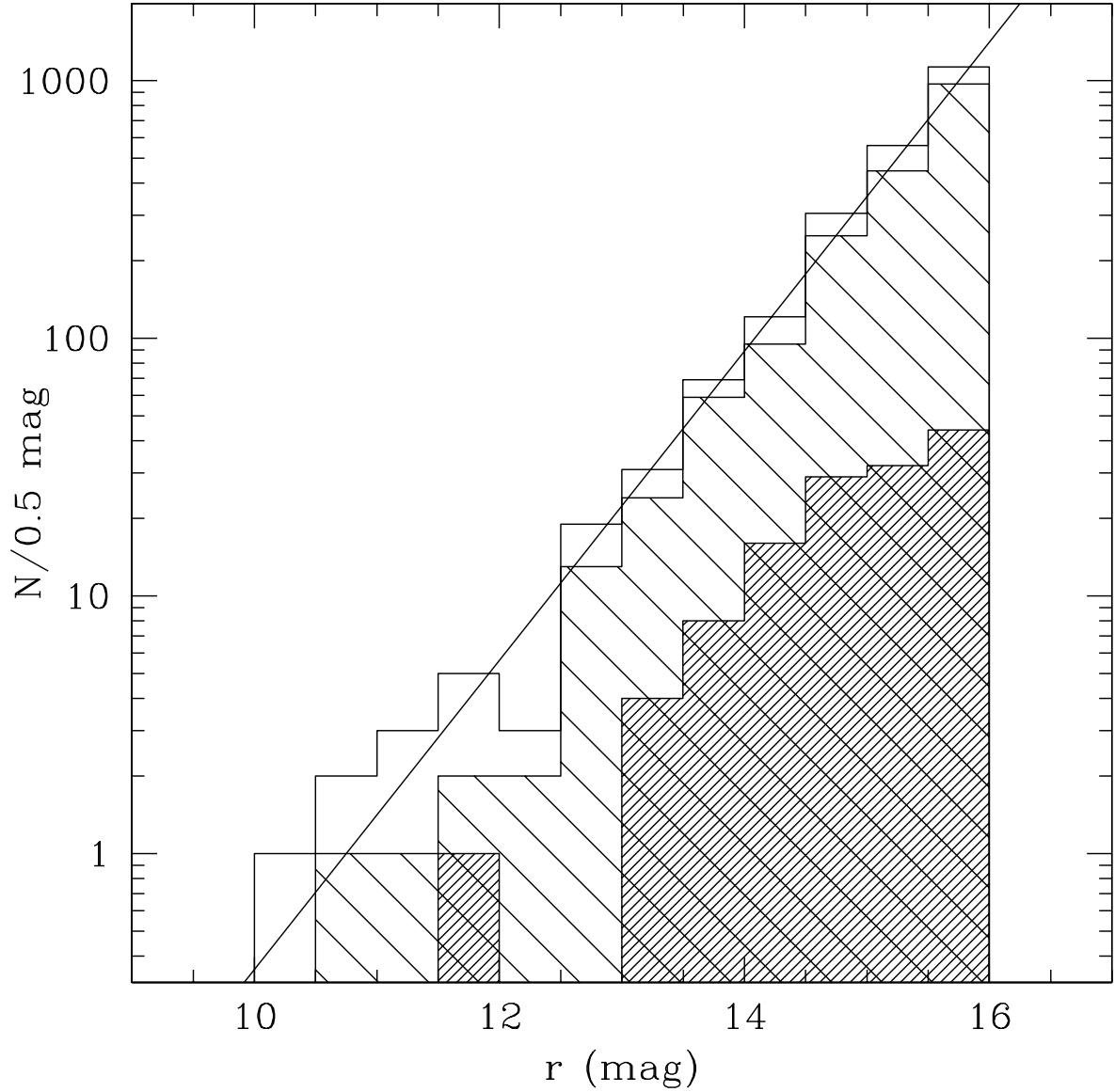}
\caption{Number of galaxies in our catalogue 
per 0.5 mag as a function of $r$ mag.
The spectroscopic sample is shown with light shading.
Galaxies that carry a flag for photometric errors are indicated by
thick shading. The curve shows the Euclidean growth, $N\sim 10^{0.6r}$.
\label{fig11}}
\end{figure}

\begin{figure}
\plotone{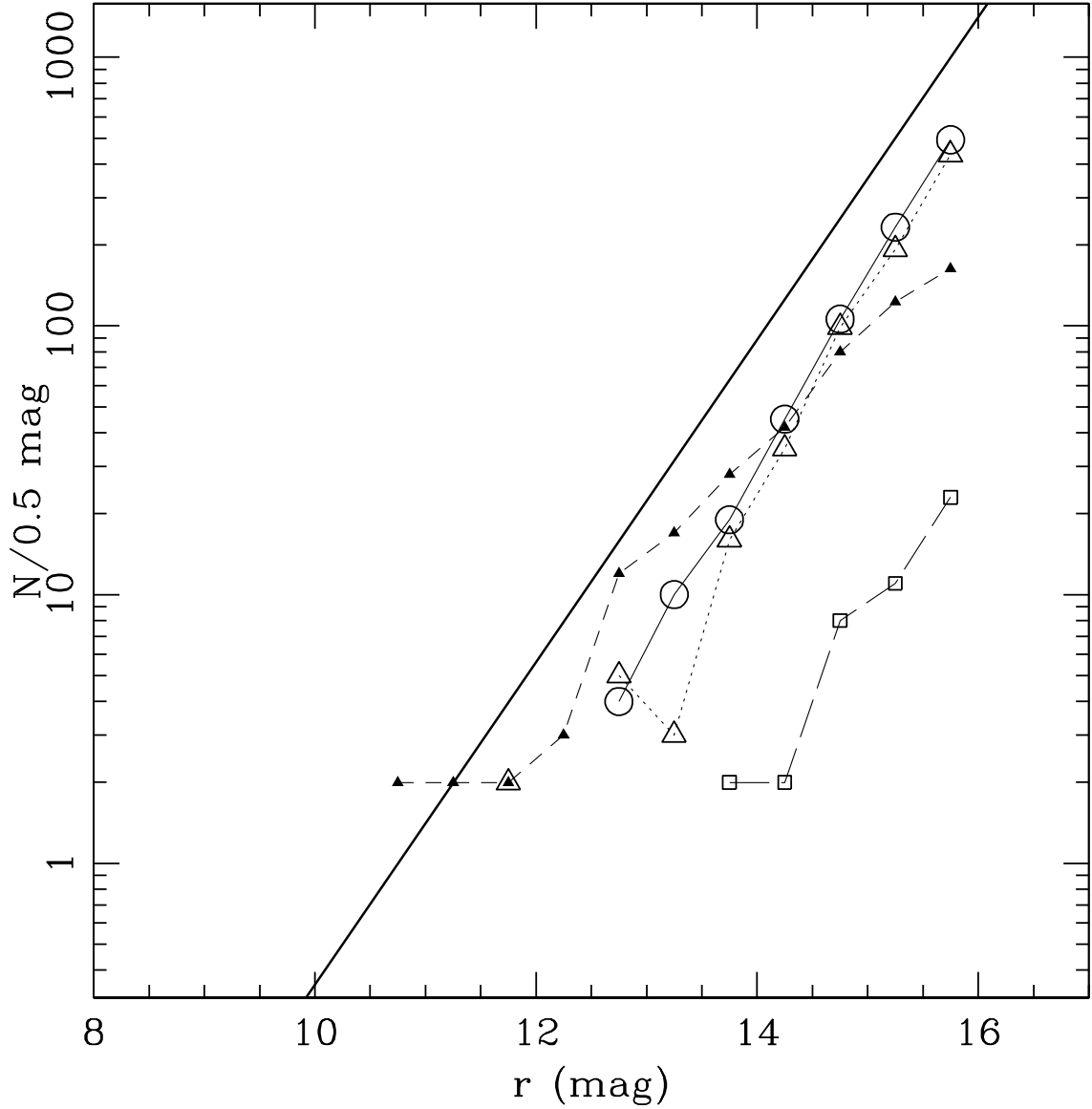}
\caption{
The same as Figure 11 but for each type. Points with $N\ge2$
are plotted. Circles, open triangles, filled triangles, squares indicate
$T=0-1$, $1.5-3$, $3.5-5$, and $5.5-6$, respectively.
The thick solid line denotes the line of $N\sim 10^{0.6r}$
shown in Figure 11.
\label{fig12}}
\end{figure}

\begin{figure}
\plotone{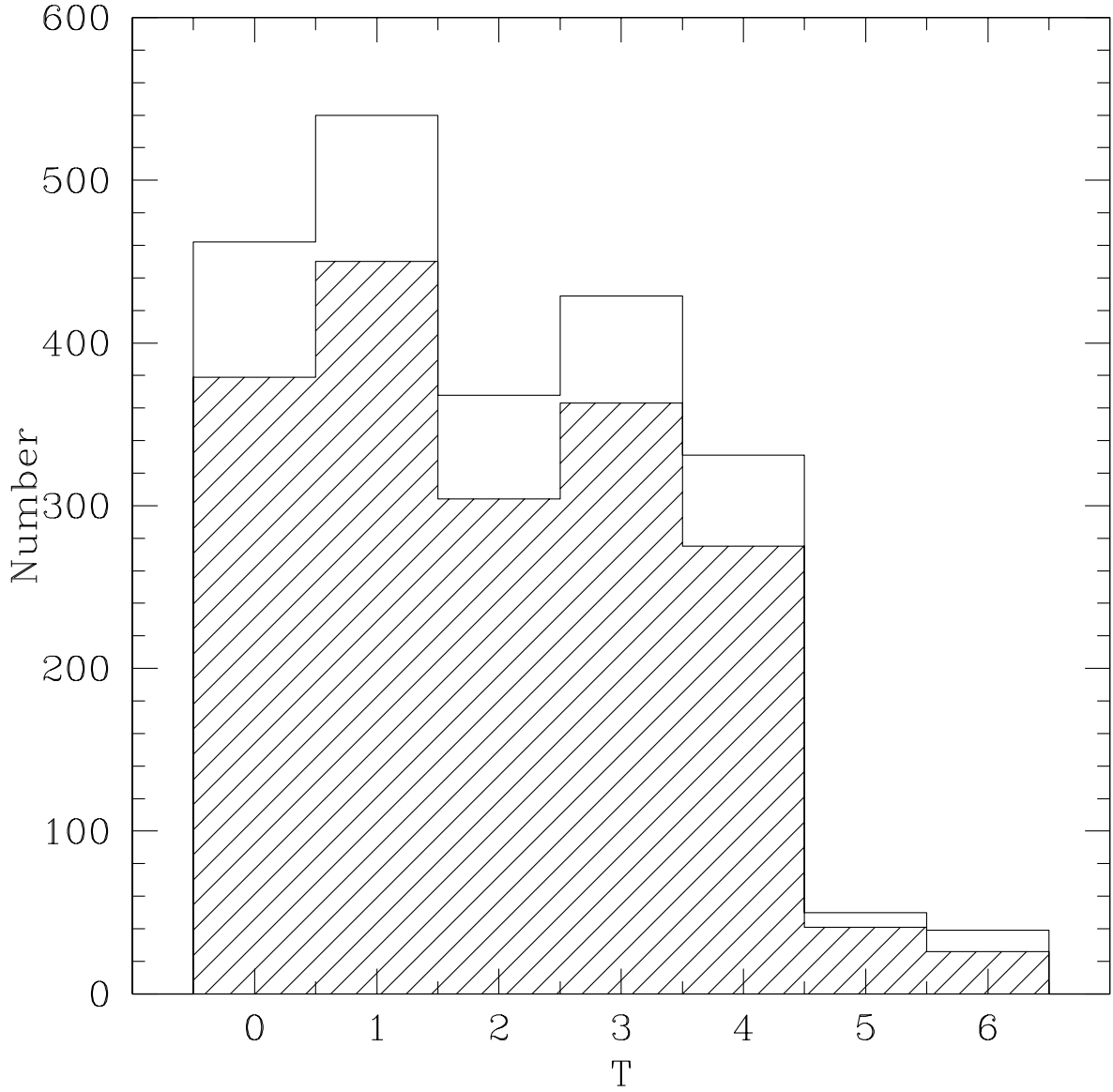}
\caption{Distribution of the morphological types in the galaxy sample
selected with $r\le 16$. Galaxies with half-integer $T$ indices are
grouped into each adjoining integer bins.  Shading represents the
spectroscopic sample. \label{fig13}}
\end{figure}

\end{document}